\input{aipcheck}

\documentclass[
    ,final            
  ]
  {aipproc}

\layoutstyle{6x9}

\begin{document}

\title{Magnetar Driven Bubbles and the Origin of Collimated Outflows from GRBs  }


\author{N.Bucciantini}{
  address={Astronomy Department, University of California at Berkeley, 601 Campbell Hall, 94720, Berkeley, CA, USA}
}

\author{E. Quataert}{
  address={Astronomy Department, University of California at Berkeley, 601 Campbell Hall, 94720, Berkeley, CA, USA}
}

\author{J. Arons}{
  address={Astronomy Department, University of California at Berkeley, 601 Campbell Hall, 94720, Berkeley, CA, USA}
}

\author{B. Metzger}{
  address={Astronomy Department, University of California at Berkeley, 601 Campbell Hall, 94720, Berkeley, CA, USA}
}

\author{T.A. Thompson}{
  address={Department of Astrophysical Sciences, Peyton Hall, Ivy Lane, Princeton, NJ, 08544, USA}
}

\begin{abstract}
 We model the interaction between the wind from a newly formed rapidly rotating 
magnetar and the surrounding progenitor. In the first few seconds after core
 collapse the magnetar inflates a bubble of plasma and magnetic fields behind
 the supernova shock, which expands asymmetrically because of the pinching
 effect of the toroidal magnetic field, as in PWNe, even if the host star is
 spherically symmetric. The degree of asymmetry depends on  the ratio of the
 magnetic energy to the total energy in the bubble. We assume that the wind 
by newly formed magnetars inflating these bubbles  is  more
 magnetized than for PWNe. We show that for a magnetic to total power 
supplied by
 the central magnetar $\sim 0.1$ the bubble expands relatively spherically while
 for values greater than 0.3, most of the pressure in the bubble is exerted 
close to the rotation axis, driving a collimated outflow out through the
 host star. This can account for the collimation inferred from observations 
of long-duration gamma-ray bursts (GRBs).
Given that the wind magnetization increases in time, we thus suggest that the
 magnetar-driven bubble initially expands relatively spherically (enhancing 
the energy of the associated supernova) while at late times it becomes
 progressively more collimated (producing the GRB). Similar processes may 
operate in more modestly rotating neutron stars to produce asymmetric 
supernovae and lower energy transients such as X-ray flashes.
\end{abstract}

\maketitle


\section{Introduction}

The increasing evidence for SN-GRB association, suggests that the
relativistic outflow associated with the burst might be related to
the formation, at the center of the progenitor, of a compact remnant,
either a black hole or a neutron star. A newly formed rapidly rotating
magnetar  \citep{usov92,thom04} or a high density accretion disk around
a central black hole \citep{macfad99} are at the moment the most
promising engines. 

If magnetars are born with millisecond period, their rotational 
energy ($\sim 10^{52}$ ergs) is larger than the energy of the associated 
supernova explosion ($\sim 10^{51}$ ergs). Given the strong magnetic 
field,  and the relatively short spin-down time, they can in principle 
release most of their energy, in the form of a magnetized wind in a few 
minutes, as required for powering a GRB. However to properly assess the 
validity of magnetar as possible sources for long duration GRBs, one must 
also verify that the conditions of the proto-magnetar outflow agree with
the observed properties of GRBs, in particular that the wind can be 
accelerated to high Lorentz factors $\gamma$ and that most of the energy 
is collimated, as observational constraints suggest \citep{rho99}.

\section{Proto-Magnetar winds}

After the core-collapse, the newly born magnetar cools and contracts on 
timescales, $\sim10-100$\,s, radiating its gravitational binding energy 
($\sim$$10^{53}$ ergs) in neutrinos of all species \citep{pons99}. This 
cooling phase is accompanied by a thermal wind, driven by neutrino energy 
deposition, which emerges into the post-supernova-shock ejecta. As the 
neutrino luminosity decreases, the thermal pressure at the proto-neutron 
star surface decreases, and the wind becomes magnetically dominated, so 
that the outflow must eventually become relativistic. If the surface
magnetic field is of order $B_0\sim 10^{15}$ G, the transition from a 
mass loaded to a magnetic dominated wind takes place just a few seconds 
after the core-bouce \citep{thom03}.

Even if observations clearly show that ultrarelativistic outflows are 
produced by pulsars, untill now theoretical models have failed to explain 
the observations. The conditions of the wind can be parametrized in 
terms of $\sigma=\Omega^2\Phi^2/\dot{M}$, where $\Omega$ is the rotation 
rate, $\Phi$ the magnetic flux and $\dot{M}$ the mass flux. $\sigma$ is 
the maximum Lorentz factor the wind can achieve if all magnetic energy 
is converted into kinetic energy, and at the base of a proto-magnatar it 
changes from values $\sim 1$ to $\sim 10^3$ in tens of seconds. The 1D 
results  \citep{michel69} show that the asymptotic Lorentz factor only 
reaches values $\sim \sigma^{1/3}$, not enought for a GRB. This is known, 
in the pulsar literature as the $\sigma-\gamma$ problem \citep{arons04}. 

Both in the case of a dipolar and monopolar magnetic field, solutions 
\citep{me06} show that for $\sigma<1$, the energy and angular momentum 
losses at large distances are directed along the rotation axis. The outflow 
velocity tends to be higher at intermediate latitudes, contrary to the 
simple expectation that magnetocentrifugal acceleration should be more 
efficient along the equator. The flow is however almost thermally driven 
and the asymptotic speed non relativistic. As soon as $\sigma$ exceeds 
$10$ the flow becomes almost radial, the energy and angular momentum 
fluxes rapidly approach the force-free values.  Even if $v \ll c$ inside the 
Light Cylinder, asymptotic results agree with the flow structure found 
by Bogovalov \citep{bog01}. Energy flux scales as $\sin^2{(\theta)}$, 
and  $\gamma\propto\sin{(\theta)}$, in agreement with the solution of 
the exact monopole. However the terminal $\gamma$ remains 
$< \sigma$, pointing to inefficient acceleration. 


\begin{figure}
  \includegraphics[height=.3\textheight]{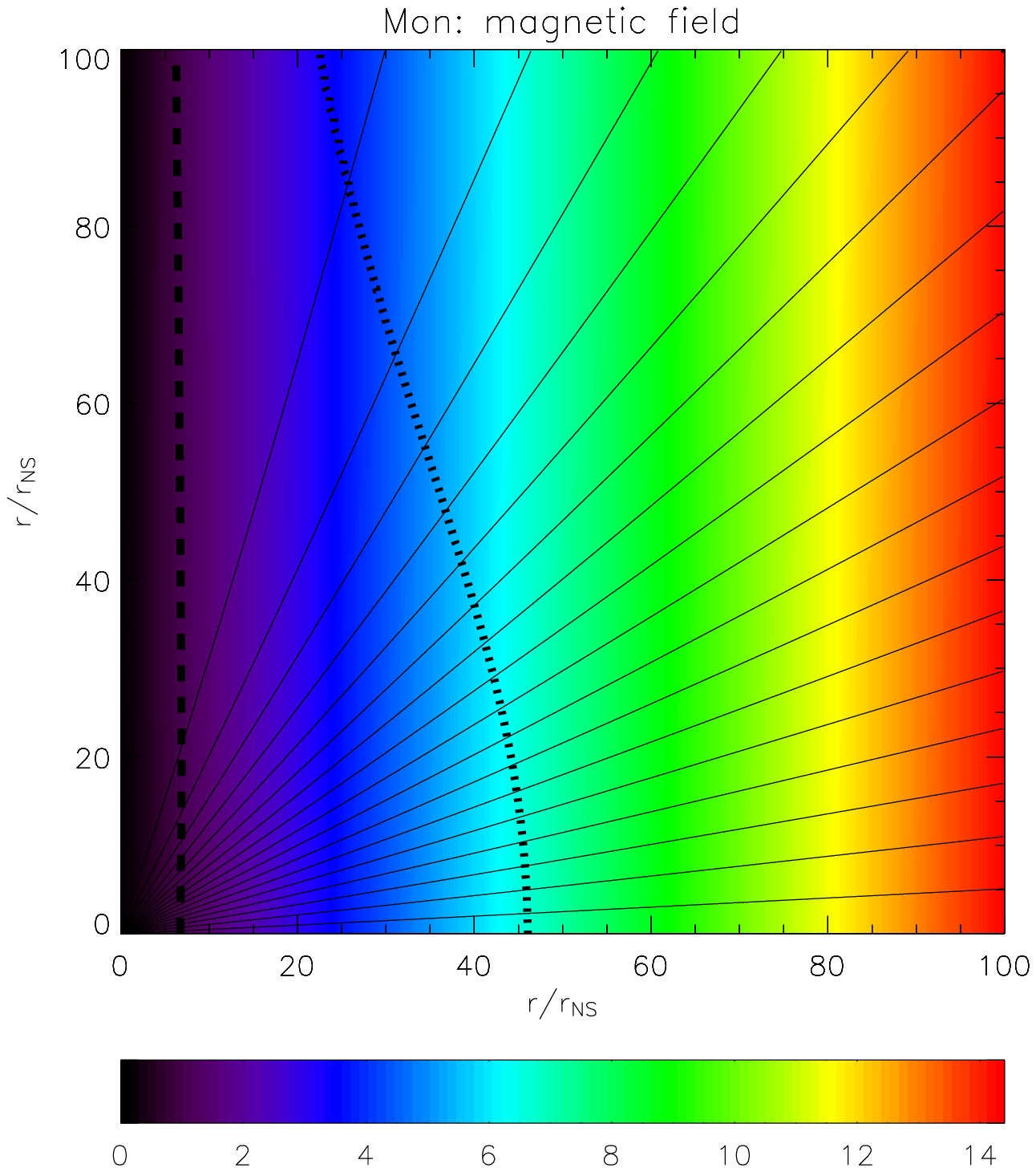}
  \includegraphics[height=.3\textheight]{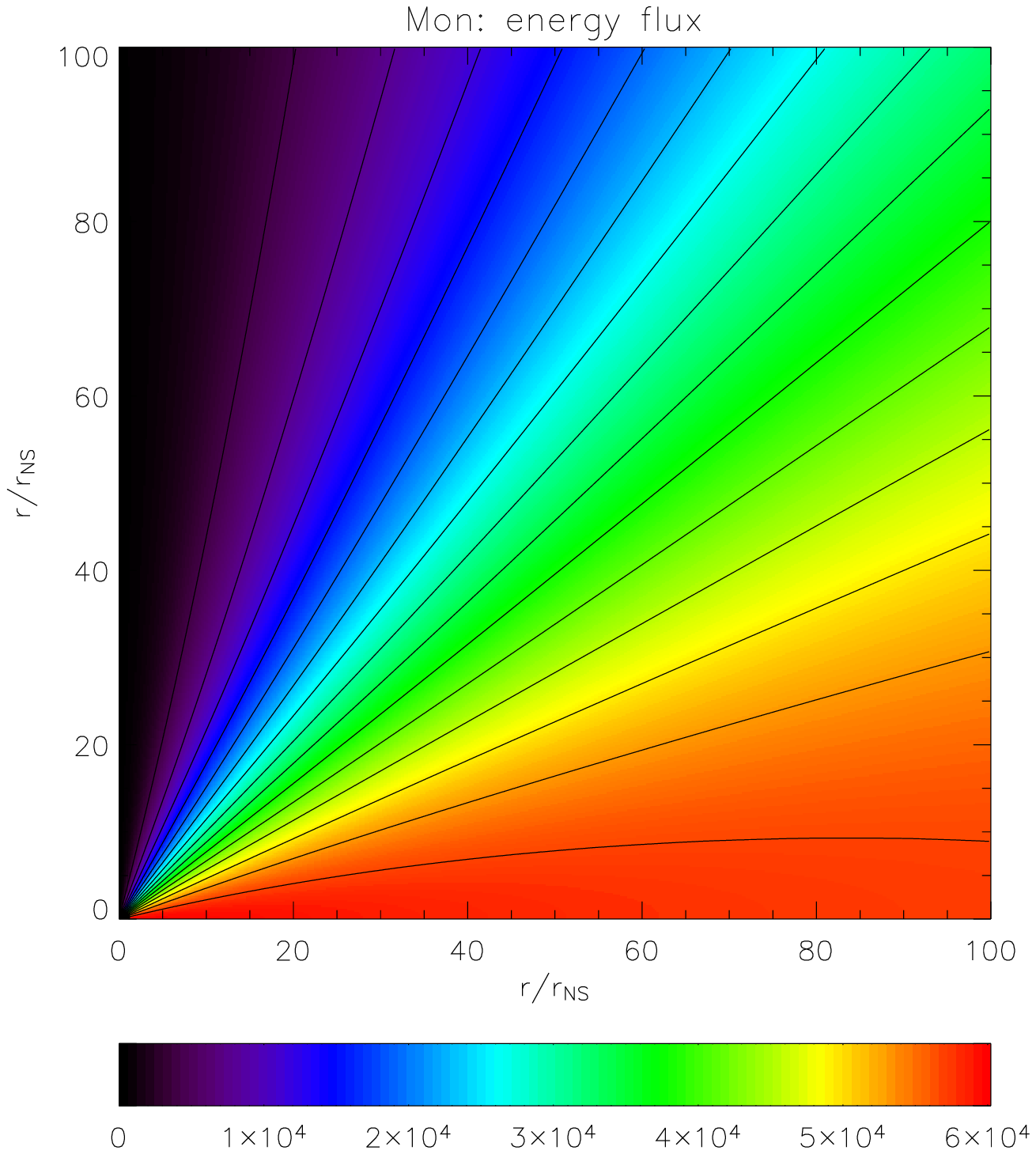}
  \caption{Structure of a proto-magnetar wind for $\sigma=200$ 
\citep{me06}. Left panel: solid contours represent the poloidal magnetic 
field, dashed line is the Light Cylinder, dotted line is the fast surface. 
Colors represent the toroidal magnetic field $\propto \sin{(\theta)}$. 
Right panell: enegy flux $\propto \sin^2{(\theta)}$. }
\label{fig:wind}
\end{figure}


If one considers the properties of the wind at a distance of $\sim10^9$ cm, 
comparable with the typical size of the collapsed core, one finds that 
for $\sigma<5$, the flow does collimate along the axis, and the energy flux 
is higher at higher latitudes, however acceleration is inefficient, and by 
the time the flow reaches equipartition the Lorentz factor is still small. 
For values of $\sigma\sim 10$, the energy distribution in the wind appears 
to be more isotropic. For values of $\sigma > 30$ the energy flux is mostly 
directed along the equatorial plane, and there is no evidence for the 
presence of a collimated jet.

By coupling numerical results for the wind energetics with theoretical models
for the evolution of the neutrino luminosity, it is possible to derive 
spin-down curves for the proto-magnetar \citep{met07}. This shows that the 
magnetar luminosity, at times longer than one second after core bounce, 
drops as $t^{-1}$, while the magnetization $\sigma$ can reach values 
$\sim 10^{2-3}$ in about 10-20 seconds.  

The numerical results about proto-magnater wind suggest that it is not 
possible to achieve simultaneously collimation and acceleration to high 
Lorentz factors. It is thus fair to conclude than the collimated 
relativistic outflow that is observed in GRBs cannot be the free-flowing 
 proto-magnetar wind. However the wind is supposed to interact with
 the heavier progenitor \citep{um06,um07}, and as we will show in the 
next section, this can provide a viable mechanism 
for collimation.

\section{Magnetar Wind Nebulae (MWNe)}

The interaction of the relativistic wind from a proto-magnetar with the 
surrounding progenitor can be modelled on the theory of Pulsar Wind Nebulae 
(formed by the interaction of a pulsar wind with the SNR). The basic 
idea is that, due to the confinement of the heavier progenitor, the 
proto-magnetar wind is slowed down in a strong termination shock, which 
heats and compresses the plasma forming a hot magnetized bubble, that we 
call MWN. The evolution and the internal dynamics in the MWN can lead 
to the formation of collimated outflows that, emerging from the star,
might give rise to jets.

Two aspect of the PWNe structure are here considered. The presence of 
predominantly toroidal magnetic field, as expected for the wind of a 
compact rotator, creates inside the bubble a pressure anisotropy, with 
a higher axial pressure that leads to the formation of elongated bubbles, 
along the rotation axis of the pulsar. Moreover X-ray observations of 
PWNe have shown the presence of a mildly relativistic collimated 
jet/plume inside the PWN \citep{weiss00}, which could in principle 
be a less energetic counterpart of GRB jets.

For proto-magnetar with millisecond rotation period the location of 
the Alfvenic surface is at $\sim 10^7$ cm, and the fast magnetosonic 
surface is at $\sim 10^7-10^8$ cm, much smaller than the typical size 
of the collapsing core of the stellar progenitor $\sim 10^9$ cm. If 
this cavity has been evacuated by the outgoing SN shock, the location 
of the termination shock is $> 10^8$ cm, so one can assume a superfast 
injection of energy and toroidal magnetic field. If one neglects the 
plasma flow inside the bubble, a simple solution for the pressure 
distribution can be obtained in the case of toroidal magnetic field 
\citep{beg92}. This solution provides an anisotropic pressure, higher 
on the axis, with a typical highscale depending on the ratio of magnetic 
to total energy in the nebula, which in turn is a function of the wind 
injection properties. 

Even if most of the energy flux is concentrated in the equatorial plane, 
given the relative size of the cavity and termination shock, the 
surrounding bubble will still reach magnetohydrostatic equilibrium, 
and will elongate along the axis. To model the evolution of the bubble 
in a simplified way one can assume it to be bounded by a thin shell of 
swept-up material, whose evolution depends on the internal pressure 
distribution, properly computed taking into account injection and 
adiabatic losses \citep{me07a}. 

For $\dot{E_{mag}}/\dot{E_{tot}}\sim 0.1$ the pressure distribution 
inside the bubble is relatively spherical and so the surrounding shell 
becomes only moderately asymmetric. Most of the energy supplied by the 
magnetar is in this case transferred to the SN shock and the host star. 
For larger values of $\dot{E_{mag}}/\dot{E_{tot}}$ the elongation 
increases because most of the pressure is exerted along the axis, and 
there is evidence for a very elongated channel being driven through the 
surrounding stellar envelope. It is also worth noting that the high 
axial pressure produced by a magnetized bubble with energy $E$ leads 
to an expansion along the axis that is quantitatively similar to that 
produced by a jet with kinetic power
\begin{equation}
L_{jet}\simeq 7.5\cdot 10^{51} \, \theta \left(\frac{E}{10^{51} {\rm
erg}}\right)\left(\frac{10^9 {\rm cm}}{R}\right) {\rm erg \, s^{-1}},
\end{equation}
where we have assumed that the angular collimation $\theta \sim H/R$ and 
that the dimensionless axial pressure $P \propto R/H$ \citep{me07a}.


\begin{figure}
  \includegraphics[height=.3\textheight, bb= 80 364 400 717, clip]{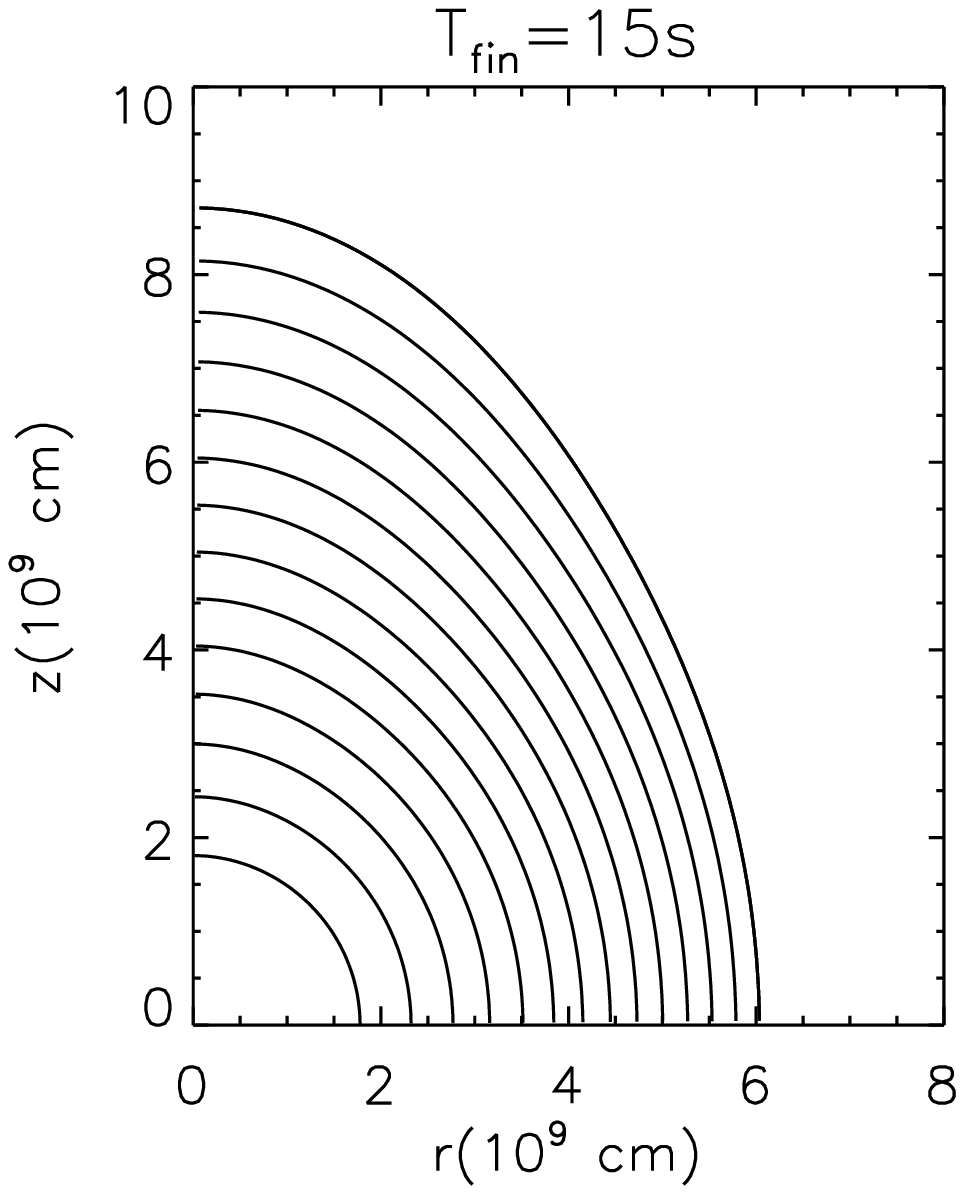}
  \includegraphics[height=.3\textheight, bb= 80 364 295 717, clip]{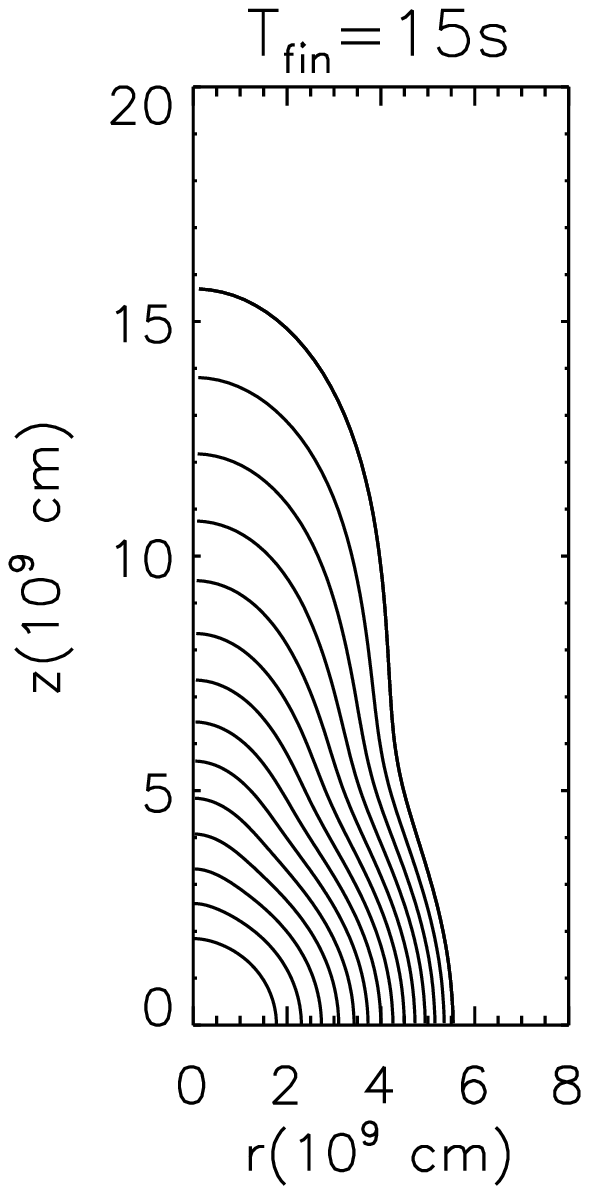}
  \includegraphics[height=.3\textheight, bb= 80 364 245 717, clip]{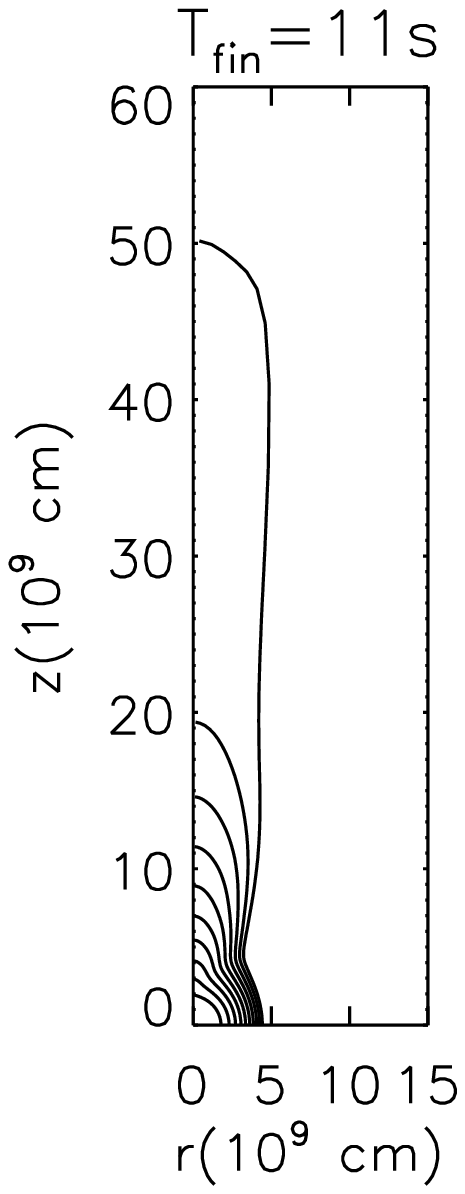}
  \caption{Evolution of a magnetar wind nebula, inside a progenitor, for 
three different values of the wind magnetization  \citep{me07a}. From left 
to right $\dot{E_{mag}}/\dot{E_{tot}}=0.1,0.2,0.3$. Contours are spaced 1 
second apart.}
\label{fig:bubble}
\end{figure}


To properly model the dynamics and evolution of the flown inside the MWN, 
to assess the validity of the thin-shell approximation and to verify if 
the presence of an internal jet/plume can lead to relativistic collimated 
jets, one must resort to numerical simulations  \citep{me07b}. 

In the post termination shock region
 the flow undergoes a large-scale circulation with much of the
matter being diverted from the equatorial region to the pole where it
flows out along the jet, just as in analogous calculations for
the structure of PWNe \citep{kom04,ldz04}.  This is caused
by the toroidal magnetic field that builds up in the MWN.  Although
collimation by hoop stresses is ineffective for the initially
relativistic flow in the free wind, the non-relativistic $v \sim 0.5
c$ flow in the post termination shock region is effectively collimated
by the magnetic field.

 At t = 4 sec, the MWN
and jets are still fully contained within the star.  By 5 sec, 
however, they have reached the
radius of the star at $\approx 2.5\,10^{10}$ cm and at 6 seconds,
the material in the jet at large radii has escaped the star and has
accelerated back up to $v \approx c$. 

\begin{figure}
  \includegraphics[height=.3\textheight,bb= 22 0 414 450, clip]{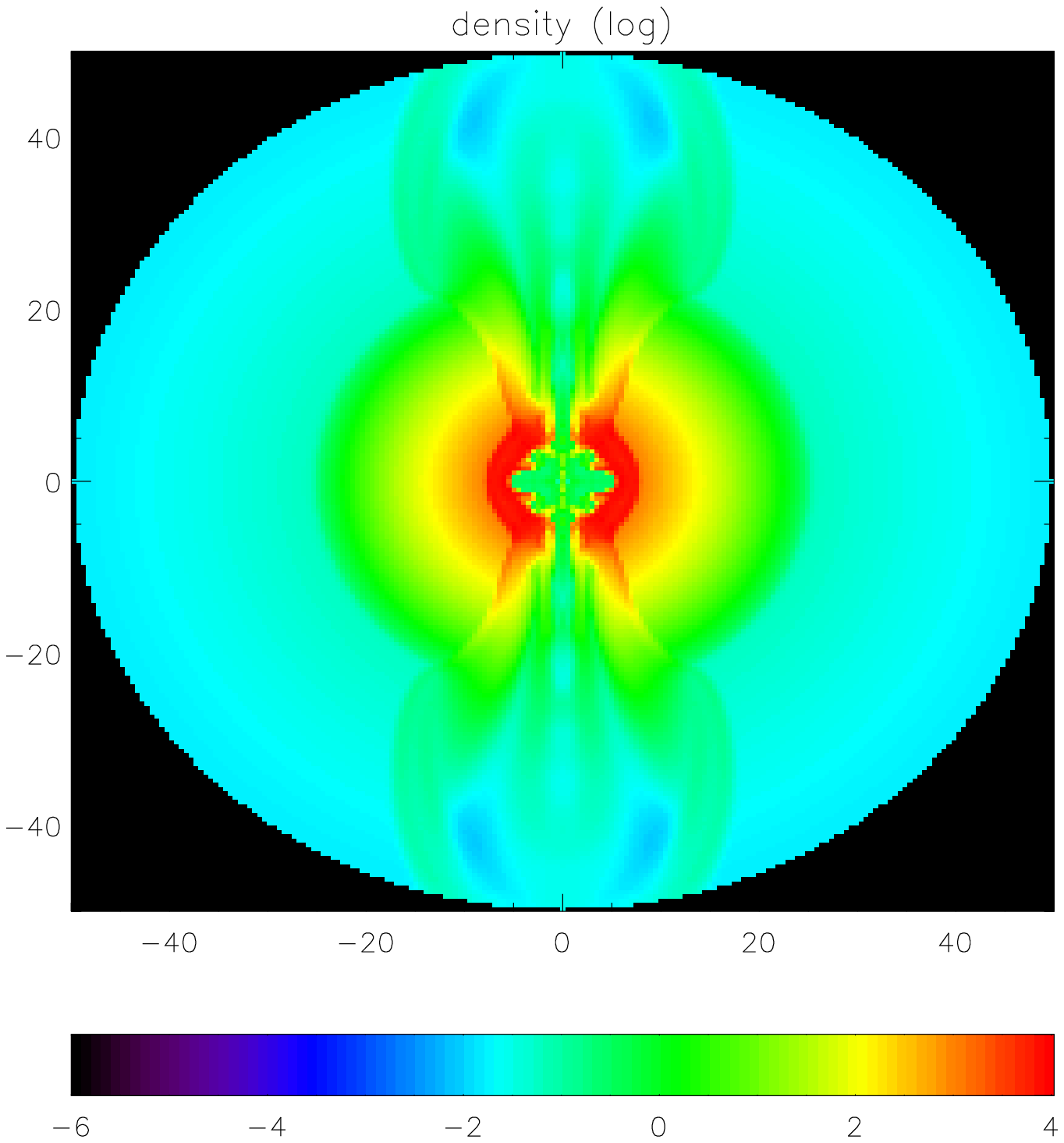}
  \includegraphics[height=.3\textheight,bb= 22 0 414 450, clip]{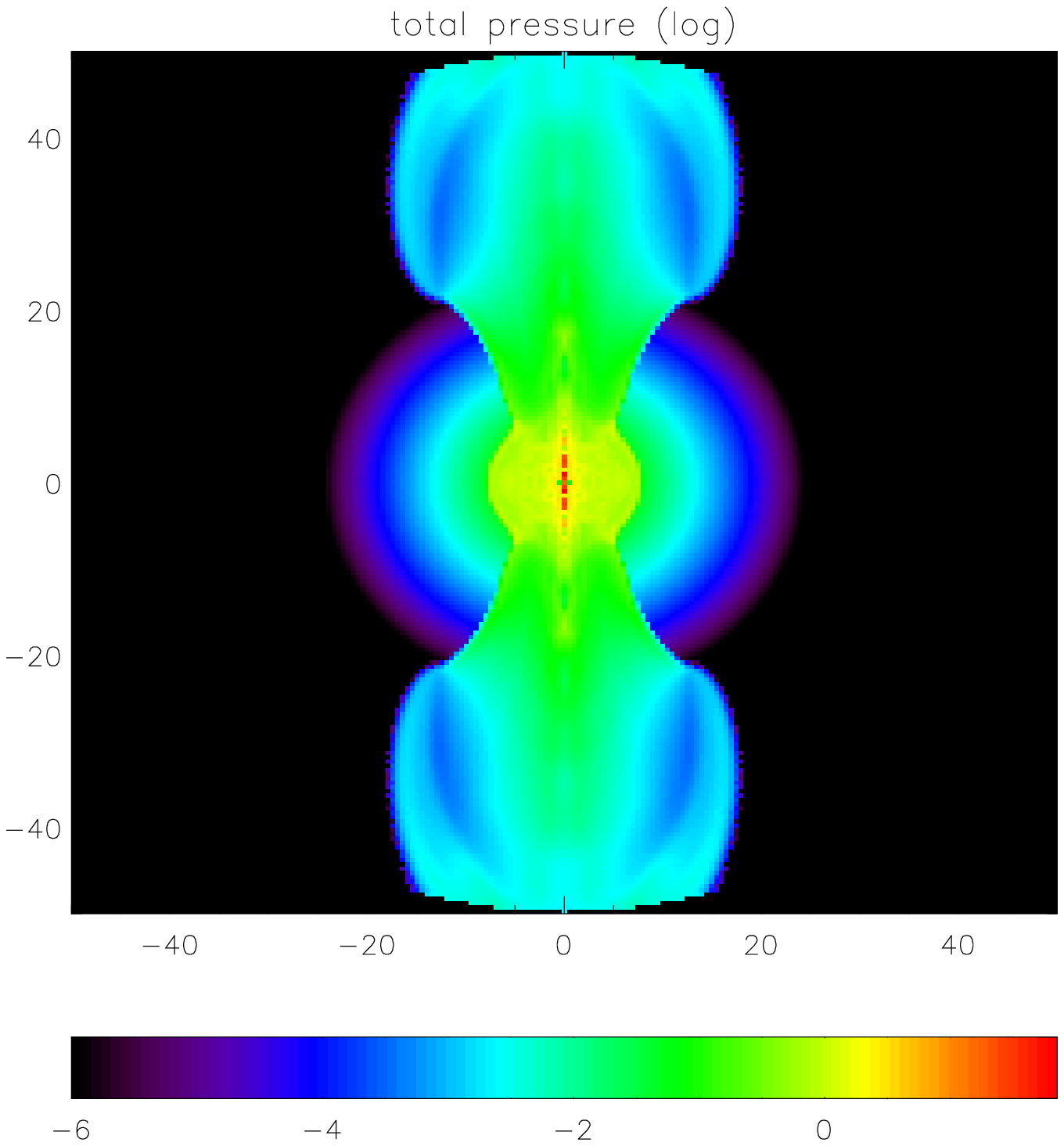}
  \includegraphics[height=.3\textheight,bb= 22 0 414 450, clip]{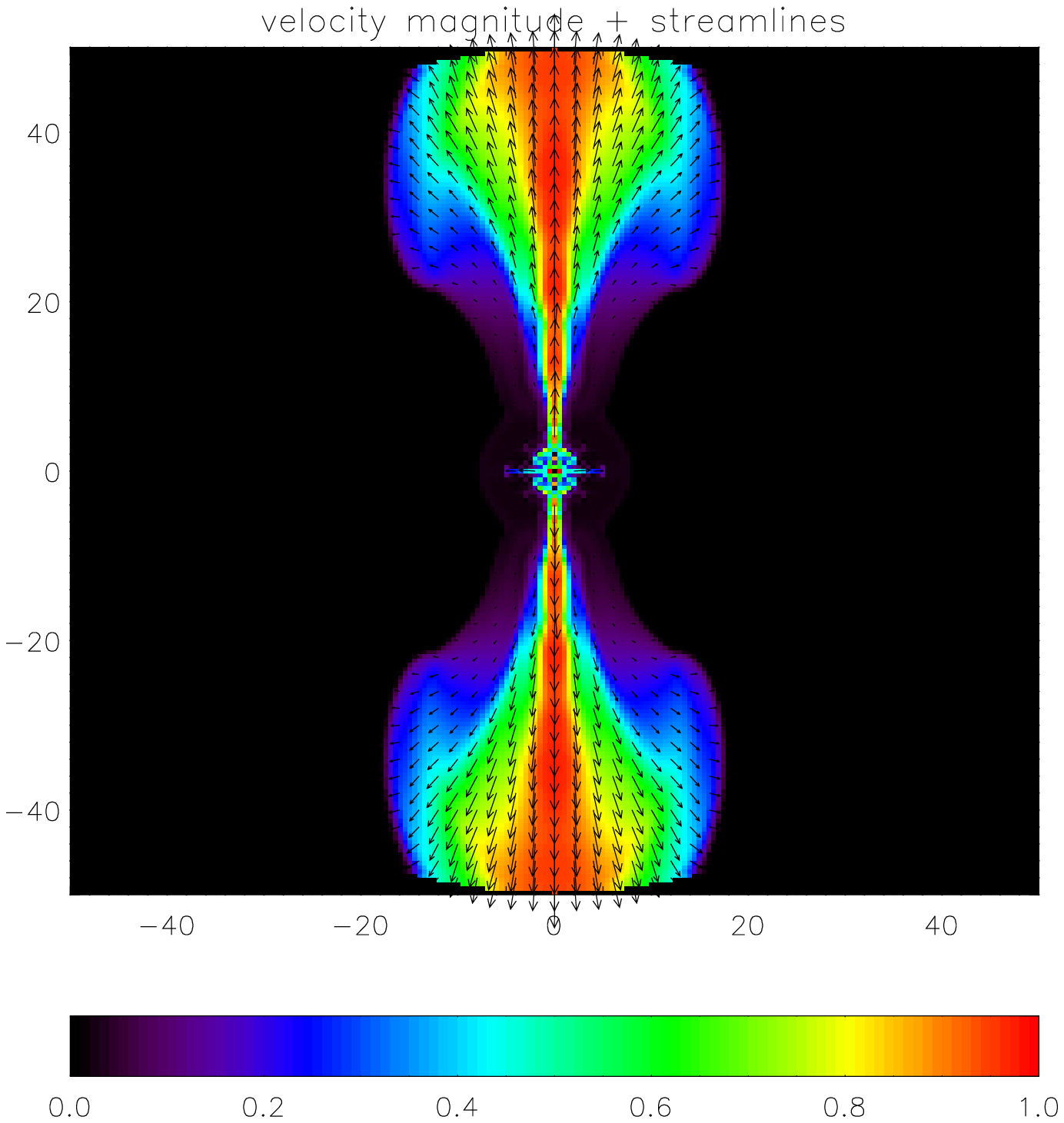}
  \caption{Relativistic MHD simulations of the interaction of a 
proto-magnetar wind with the host star \citep{me07b}. From left to 
right: density, pressure and velocity after 6 second. Note the formation 
of a collimated relativistic jet with opening angle $\sim 5^\circ$.}
\label{fig:grb}
\end{figure}

Although the magnetic field is crucial for generating and collimating
the outflow seen in Figure \ref{fig:grb}, the field is not
energetically dominant; the ratio of the magnetic energy to the
thermal energy in the bubble is typically $\sim 0.1-0.2$, although it
can reach $\sim 1$ in the region where the hoop stress is most
effective.  The enthalpy of the shocked gas in
the post-termination-shock region primarily determines the asymptotic
Lorentz factor in the jet.  In turn, the enthalpy of the shocked gas
is determined by the wind Lorentz $\gamma$ and $\sigma$.
 The acceleration is roughly linear in radius,
consistent with that expected in a ``fireball'' with a constant
opening angle. At late times nearly all of the energy
supplied by the central magnetar escapes to large radii and a
significant fraction of the energy is carried by the high velocity
core of the jet.


The jet escaping the star at large
distances has fluctuations in $\gamma$ and $\dot E$ on $\sim 0.01-0.1$
second timescales.  It is natural to
speculate that this variability in the source region could manifest
itself as variability in the gamma-ray emission in GRBs.

We also find that within our model there is no significant transfer of energy 
from the MWN to the SN shock at intermediate latitudes, which suggests 
that it is unlikely that this mechanism could be responsible both for a 
GRB and an hyper-energetic SN, unless the earlier low-magnatization wind may
contribute significant energy to the SN shock.
 
\section{Conclusions}

Our simulations demonstrate that a spinning down magnetar inside a
massive star can divert most of its spin-down energy to an axial jet,
even in the conservative case in which the host star and magnetar wind
are spherically symmetric.  Physically, this occurs because of the
strong toroidal magnetic field in the bubble of relativistic plasma
that is at first inertially confined by the
progenitor star and SN shock.  The magnetic field creates an
asymmetric pressure distribution in the bubble, with the pressure much
larger at the pole than at the equator.  In addition, the toroidal field
collimates the moderately relativistic flow behind the termination
shock into an axial jet. This collimated
outflow would thus have physical properties similar to those required
to produce GRBs.


\begin{theacknowledgments}
N.B. was supported by NASA through Hubble Fellowship grant
HST-HF-01193.01-A, awarded by the Space Telescope Science Institute,
which is operated by the Association of Universities for Research in
Astronomy, Inc., for NASA, under contract NAS 5-26555.
\end{theacknowledgments}





\IfFileExists{\jobname.bbl}{}
 {\typeout{}
  \typeout{******************************************}
  \typeout{** Please run "bibtex \jobname" to optain}
  \typeout{** the bibliography and then re-run LaTeX}
  \typeout{** twice to fix the references!}
  \typeout{******************************************}
  \typeout{}
 }



\end{document}